# Towards a theory of flow stress in multimodal polycrystalline aggregates. Effects of dispersion hardening

D. Cevizovic[1,a)], A.A. Reshetnyak[2,b)], Yu.P. Sharkeev[2,3c)]

[1]*University of Belgrade, "Vinca" Institute of Nuclear sciences, Belgrade, Serbia*
[2]*Institute of Strength Physics and Materials Science SB RAS, 2/4, Ave. Akademicheskii, Tomsk, 634055, Russia*
[3]*National Research Tomsk Polytechnic University, 10, Ave. Lenin, Tomsk, 634011, Russia*

[b)] Corresponding author: reshet@ispms.tsc.ru
[a)] cevizd@vin.bg.ac.rs, [c)] Sharkeev@ispms.tsc.ru

**Abstract** We elaborate the recently introduced theory of flow stress, including yield strength, in polycrystalline materials under quasi-static plastic deformations, thereby extending the case of single-mode aggregates to multimodal ones in the framework of a two-phase model which is characterized by the presence of crystalline and grain-boundary phases. Both analytic and graphic forms of the generalized Hall–Petch relations are obtained for multimodal samples with BCC (α-phase Fe), FCC (Cu, Al, Ni) and HCP (α-Ti, Zr) crystalline lattices at T=300K with different values of the grain-boundary (second) phase. The case of dispersion hardening due to a natural incorporation into the model of a third phase including additional particles of doping materials is considered. The maximum of yield strength and the respective extremal grain size of samples are shifted by changing both the input from different grain modes and the values at the second and third phases. We study the influence of multimodality and dispersion hardening on the temperature-dimensional effect for yield strength within the range of 150–350K.

## INTRODUCTION

The theory [1,2,3] of flow stress in polycrystalline (PC) materials under quasi-static plastic deformations (PD) in the case of tensile strain depending on the average size of crystallites (grains) $d$ in the range of $10^{-8} - 10^{-2}$ m is based on a statistical model of mechanical energy distribution in each crystallite of a single-modal isotropic PC material with respect to quantized quasi-stationary levels. The highest energy level is chosen equal to the energy of the maximal straight-line dislocation within a disclination-dislocation deformation mechanism based on the origin of a dislocation due to certain thermal-fluctuation causes beyond the elastic limit under loading, initiated by the appearance of 0D-defect sequences (nanopores, bi-nanopores, ..., $n$-nanopores, being the zones of localized deformation and vacancies) and finalized by the formation of a dislocation itself as the ends of the above sequence reach the grain boundary (GB) or the axes of other dislocations inside the grain. The generation process for a dislocation, when regarded as an elementary act of plastic deformation in each grain, allows one to give a general description [1,2,3,4] of a non-equilibrium deformation process as a sequence of equilibrium deformation processes, taking into account the smallness of relaxation time for the atoms of a grain crystal lattice (CL) at new ground states, $\tau = a/v_s = (0.3 \ast 10^{-9})/10^3 \sim 10^{-12}$ sec, with the interatomic distance $a$ and $v_s$ being the crystallite speed of sound, in comparison with the minimal time between the acts of PD, $\Delta t_0 = a/\dot{\varepsilon}d \sim 2.47 \ast 10^{-1}$ sec, under the strain rate $\dot{\varepsilon} = 10^{-5}$ sec$^{-1}$, $(a,d) = (3 \ast 10^{-10}, 10^{-3})$ m [1,4]. The energy spectrum of each crystallite (located far from the grain boundary) consists of equal-distant energy zones (implementing the most probable assembly of dislocations with a unique Burgers vector, $b$, starting from the zero-level energy of a crystallite without defects, $E_0$, up to the level with the maximal number of atoms in a full dislocation and elementary segments in the case of partial dislocations) situated on a dislocation axis, $E_N$, $N=[d/b]$. The portion (quantum) of energy necessary for transition from a crystallite state to a neighboring state equals to the energy of a unit dislocation, ½ $Gb^3$, being almost precisely commensurate with the activation energy of an atom in a material during the diffusion process [1]. Following the corpuscular-wave hypothesis of L. de Broglie, we propose for this quantum a quasiparticle interpretation [2] under the provisional name of a *dislocon*, which proves to be extremely useful in a theoretical description of localized deformation phenomena such as the Chernov–Luders macroband or Portevin–Le Chatelier effect.

Inside the equilibrium segments for a given strain, $\varepsilon$, the emergence probabilities for possible defects during an elementary act of PD at a time instant $t=\varepsilon/\dot{\varepsilon}$ are distributed (by the Large Numbers Law) according to Boltzmann [1,3,4], namely, $P_n(\varepsilon) \sim A(\varepsilon) exp\left\{-\frac{n}{2} G b_\varepsilon^3 / NkT\right\}$, with an effective Burgers vector $b_\varepsilon = b(1+\varepsilon)$, $b_\varepsilon(T') = b_\varepsilon(T)\{1+\alpha_d(T'-T)\}$, and a shear modulus $G(T')=G(T)\{1-\alpha_G(T'-T)\}$, where the linear temperature coefficients for the linear expansion, $\alpha_d$, and shear modulus, $\alpha_G$, with $T' > T$, vary between different temperature ranges at a given phase occurring in a material (within linear approximation of $G(T)$, $b_\varepsilon(T)$ as the functions of $T$).

The distribution of scalar dislocation density $\rho$ in each crystallite, with accuracy up to the terms linear in $\varepsilon$ [1],

$$\rho(b_\varepsilon, d, T) = M(0)\frac{6\sqrt{2}}{\pi}\frac{m_0}{d^2}\varepsilon\left(e^{M(\varepsilon)b/d} - 1\right)^{-1} + o(\varepsilon^2), \quad M(\varepsilon) = Gb_\varepsilon^3/2kT, \tag{1}$$

for the energy scale $M(\varepsilon) \sim 10^2$ (inverse of the speed sensitivity), when $T \sim 300K$, and the *polyhedral parameter* $m_0 \sim 10^1 - 10^2$, leads to a flow stress $\sigma(\varepsilon)$ having the form

$$\sigma(\varepsilon) = \sigma_0(\varepsilon) + \alpha m \frac{Gb}{d}\sqrt{\frac{6\sqrt{2}}{\pi}m_0\varepsilon M(0)}\left(e^{M(\varepsilon)\left[\frac{b}{d}\right]} - 1\right)^{-\frac{1}{2}}, \tag{2}$$

according to the Taylor strain hardening mechanism, $\tau = \tau_f + \alpha Gb\sqrt{\rho}$, $\sigma(\varepsilon) = m\tau$, for $m = 3.05$, which includes at $\varepsilon = 0.002$ the normal [5] and abnormal (e.g., see [6,7] and references therein) Hall–Petch (HP) relations for coarse-grained (CG) and nano-crystalline (NC) grains. Notice that at the purely crystalline phase for a single-mode PC aggregate the value of $\sigma(\varepsilon)$ reaches its flow-stress maximum at an extreme grain of average value $d_0$ [1],

$$d_0(\varepsilon, T) = b\frac{Gb^3(1+\varepsilon)^3}{2 \cdot 1{,}59363 \cdot kT}, \tag{3}$$

estimated at $10^{-8} - 10^{-7}$ m and shifted to the region of larger grains with a decrease in temperature and increase in strain $\varepsilon$, thus determining the *temperature-dimensional effect* (TDE). Other properties of TDE consist in the fact that the example of $T$-behavior with $\sigma(0.002) = \sigma_y$ results from a cubic approximation to the HP relation implied by (2) with the HP coefficient $k(\varepsilon) = \alpha m G\sqrt{\frac{6\sqrt{2}}{\pi}m_0\varepsilon b\,M(0)/M(\varepsilon)}$ determined from experiment,

$$\sigma(\varepsilon)|_{d \gg b} = \sigma_0(\varepsilon) + k(\varepsilon)d^{-1/2}(1 - \tfrac{1}{4}M(\varepsilon)b/d) \Rightarrow d_1 = \frac{bM(\varepsilon)}{4\alpha_G}\left(\frac{1}{T} + \alpha_G\right). \tag{4}$$

The value of $\sigma(\varepsilon)$ increases as $T$ decreases in all the grains (more than $d_1 \sim 3d_0$) and then decreases at the NC region [2,4] in a purely crystalline single-mode PC aggregate [2], as shown explicitly for Al. A natural incorporation into the one-phase model [2,3,4] of an unhardening GB phase transforms it into a two-phase model with integral flow stress:

$$\sigma_\Sigma(\varepsilon) = \left(1 - n\tfrac{b}{d}\right)\sigma_C(\varepsilon, d) + (n-m)\tfrac{b}{d}\sigma_{GB}(\varepsilon, d_{GB}) - m\tfrac{b}{d}\sigma_P(\varepsilon, d_P), \quad m \leq n \tag{5}$$

with $\sigma_C(\varepsilon, d) = \sigma(\varepsilon, d)$ being the flow stress at the 1st (solid) phase in a sample having the basic grains of diameter $d$, for $\sigma_{GB}(\varepsilon, d_{GB})$ and $\sigma_P(\varepsilon, d_P)$[1] being the flow stresses having the same dependence as $\sigma_C(\varepsilon, d)$, albeit for grains and pores in the GB region having the respective average sizes $d_{GB}$ and $d_P$, with a certain constant $n \sim 10^0 - 10^2$, taking into account the average distance between the grains and having a strong dependence on the preparation of GB states. For $n = m$, all the regions at the 2nd phase are filled with pores? In general, of various diameters. The contribution of the latter to $\sigma_\Sigma(\varepsilon)$ is conditionally negative and corresponds to an "anti-crystalline" behavior of pores, with respect to which one implements, in sub-microcrystalline (SMC) and NC samples, a slippage of (group) grains. The value of $m_0$ in (2) is determined for CG materials from the limit of the normal HP law ($d \gg b$) at $\varepsilon = 0.002$, using a relation with an experimental value of the HP coefficient $k(\varepsilon)$, $m_0 = \frac{\pi}{6\sqrt{2}}\frac{k^2(\varepsilon)}{(\alpha m G)^2 \varepsilon b}\frac{M(\varepsilon)}{M_0}$, within the one-phase model approximation. Theoretical graphic HP dependence for single-mode PC aggregates, α-Fe (BCC CL); Cu, Al, Ni (FCC CL); α-Ti, Zr (HCP CL), has been presented for one- and two-phase models with different GB in [2] and [3,4], respectively, excluding the second-phase grains ($m=n$ in (5)) and including the study of TDE in both models (for Al). A good coincidence is found between the experimental and theoretical behavior of $\sigma(0.002, d)$, with a correct prediction of the extremal value of $d_0(0.002, T)$ at $T = 300$ K. TDE for a small- and large-angle GB in a two-phase model survives with respect to the predicted $T$-behavior for $\sigma_\Sigma(0.002)$; however, this effect disappears for constant pores $\bar{d}_P \approx d_0(0.002, T)$, with $\sigma_\Sigma(0.002)$ and the extremal size of grains shifted to the SMC region.

In this paper, we continue our previous study of theoretical HP relations in PC materials, α-Fe, Cu, Al, Ni, α-Ti, Zr [1,2,3,4], for dispersion hardening by incorporating the third-phase term $\sigma_{dis}(\varepsilon_{dis}, d_{dis})$, with $d_{dis}$ being the average linear size of particles (from other compounds) at another GB phase for small- and large-angle grain boundaries, as well as for constant average diameter of pores, $d_P = \bar{d}_P$, in the presence of non-vanishing second-phase grains determining a two-mode model with dispersion hardening within the entire admissible range of grains, $d$, at the crystallite phase with new values for the extreme grains and maxima, $d_{\Sigma dis0}$, $\sigma_{\Sigma dism}$. We also investigate a modification of temperature dependence in the HP law for Al within the above implementation of the weak and dispersion phases in a PC aggregate.

---

[1] In [2,4], we restrict ourselves by a two-phase model, where the crystalline, $\varepsilon_C$, porous, $\varepsilon_P$, and second-phase, $\varepsilon_{GB}$, values of strain coincide with one another, i.e., all the phases of these PC materials transform in a coherent manner as non-interacting objects. In general, PD at each phase is heterogeneous, so that in addition to the integral (flow) stress $\sigma_\Sigma(\varepsilon)$ (5), $\sigma_\Sigma(\varepsilon) = f_1\sigma_C(\varepsilon_C) + f_2\sigma_{GB}(\varepsilon_{GB}) - f_3\sigma_P(\varepsilon_P)$, for the volume parts, $(f_1, f_2, f_3) = \left(1 - n\tfrac{b}{d}, (n-m)\tfrac{b}{d}, m\tfrac{b}{d}\right)$. According to the model of equal stress, the integral strain $\varepsilon$ should be determined empirically in the same manner as $\varepsilon = f_1\varepsilon_C + f_2\varepsilon_{GB} + f_3\varepsilon_P$, according to the model of equal strain, when all compounds reach their own values of yield strength, as in the case of steel with a ferrite-perlite structure [8, 9].

# HALL–PETCH LAW FOR α-Fe, Cu, Al, Ni, α-Ti, Zr WITH THIRD-PHASE

First, one should determine the flow stress $\sigma_\Sigma(\varepsilon)$ (5), with account taken for Footnote 1, naturally extended to dispersion hardening by particles of another material:

$$\sigma_{\Sigma dis}(\varepsilon) = (1 - U_{dis})\{f_1\sigma_C(\varepsilon_C) + f_2\sigma_{GB}(\varepsilon_{GB}) - f_3\sigma_P(\varepsilon_P)\} + U_{dis}\sigma_{dis}(\varepsilon_{dis}, d_{dis})$$
$$\varepsilon = (1 - U_{dis})(f_1\varepsilon_C + f_2\varepsilon_{GB} + f_3\varepsilon_P) + U_{dis}\varepsilon_{dis}, \quad \sum_i f_i = 1, \quad 0 \leq U_{dis} \ll 1 \quad (6)$$

or, equivalently, $\sigma_{\Sigma dis}(\varepsilon) = (1 - U_{dis})\sigma_\Sigma(\varepsilon_\Sigma) + U_{dis}\sigma_{dis}(\varepsilon_{dis}, d_{dis})$, by the "third-phase" term $\sigma_{dis}(\varepsilon_{dis}, d_{dis})$, with the weight $U_{dis}$ and with $d_{dis}$ being the average linear size of particles implementing this hardening process. Here, the role of $\sigma_{dis}$ is similar to that of $\sigma_{GB}(\varepsilon_{GB}, d_{GB})$ for the second-phase grains, albeit with the appropriate shear modulus and Burgers vector $G_{dis}, b_{dis}$ now distributed either inside the first-phase grains or in the GB regions. For CG and other PC samples with $d_{dis} \ll d$, the particles may provide an increase in the value of the integral $\sigma_{\Sigma dis}(\varepsilon)$, in particular, the yield strength for $Gb^3 < G_{dis}b_{dis}^3$. However, hardening in NC samples may occur under more involved conditions, $d_0 < d_{dis} < d$, and, presumably, also $Gb^3 < G_{dis}b_{dis}^3$, with a possible unhardening for $d_{dis} < d_0 < d$ and a certain relation among unit dislocation energies, $\frac{1}{2}Gb^3, \frac{1}{2}G_{dis}b_{dis}^3$, following the results of [1,2]. One may distinguish the cases of coherent (according to the Mott–Nabarro approach [9]), semi-coherent, and non-coherent incorporation of particles into a matrix composed of the two-phase model with basic first-phase grains.

We consider the case of a uniform distribution of third-phase Cu-particles whish possess a polyhedral form with a fixed average diameter $d_{dis}=1.5d_0$ (Cu) and are situated in the second (weak)-phase region (namely, GB) with a constant weight U$_{dis}$=0.01. For simplicity, we suppose that the strain value $\varepsilon = 0.002$ for integral yield strength, $\sigma_{\Sigma disy}$ is determined by the strain value $\varepsilon_C = 0.002$ with $\sigma_y$ at the first (crystalline) phase. To determine the values of the constant $m_0=m_0(k(\varepsilon))$ in the two-phase model, we use the known experimental values of the HP coefficient $k(0,002)$ for PC single-mode samples with BCC, FCC and HCP CL from Table 1 with small-angle GBs, corresponding to the values of $\sigma_0$, $G$, the lattice constant $a$ [10], the Burgers vectors of the least possible length $b$ and the respective most realistic sliding systems (see Table 2 [2]), the interaction constant for dislocations $\alpha$ [6,7] and the calculated values of least unit dislocations $E_d^{L_e}$, extreme grain sizes $d_0$, maximal differences of yield strength $\Delta\sigma_m$, $\Delta\sigma_{\Sigma dism}$, in accordance with (5) in [1] and (5), (6) at $T$=300 K.

**TABLE 1**: The values of $\sigma_0$, $\Delta\sigma_m =(\sigma_m-\sigma_0)$, $\Delta\sigma_{\Sigma dism}=(\sigma_{\Sigma dism}-\sigma_0)$, $E_d^{L_e}$, $k$, $m_0$, $\alpha$ in BCC, FCC and HCP polycrystalline metal samples with $d_0$, $b$, $G$ obtained using the data of [1,4,6] at $\varepsilon = 0,002$ and $d_{\Sigma dis0}$ obtained using Fig. 1, with $d_{dis}=1.5d_0$(Cu)=21.6 nm, $(d_{GBs}, d_{GBg})=(d_{Ps}, d_{Pg}) = (0.025, 0.15)*d$, correspond to the average weights of the phases: the constants U$_{dis}$=0.01 and 0.99*(1-$nb/d$;($n-m)b/d$;$mb/d$) are equal to (0.942; 0.024; 0.024) at small angles and (0.762; 0.114; 0.114) at large angles, whereas $\bar{d}_{GB}$=0.15*$d$ with constant size of the porous $\bar{d}_P = d_0$ for each PC samples, when it exists. The lowest boundary $d_{LB}$ admissible for the existence of samples with $\bar{d}_P$ is estimated as $d_{LB}$(α-Fe; Cu; Al; Ni; α-Ti; Zr) ≈(18;12;12;18;18;18) nm, which is less than in the case of respective two-phase models with $d_{dis} = d_{GB} = 0$ [4].

| Type of CL | BCC | FCC | | | HCP | |
|---|---|---|---|---|---|---|
| Material | α-Fe | Cu | Al | Ni | α-Ti | Zr |
| $\sigma_0$, MPa | 170 (anneal.) | 70 (anneal.); 380 (cold-worked) | 22 (anneal. 99,95%); 30 (99,5%) | 80 (anneal.) | 100(~100%); 300 (99,6%) | 80-115 |
| $b$, nm | $\frac{\sqrt{3}}{2}a$=0.248 | $a/\sqrt{2}$=0.256 | =0.286 | $a/\sqrt{2}$=0.249 | $a$=0.295 | $a$=0.323 |
| $G$, GPa | 82.5 | 44 | 26.5 | 76 | 41.4 | 34 |
| $T$, K | 300 | 300 | 300 | 300 | 300 | 300 |
| $k$, MPa·m$^{1/2}$ | 0.55-0.65 ($10^{-5}-10^{-3}m$) | 0.25 ($10^{-4}-10^{-3}m$) | 0.15 ($10^{-4}-10^{-3}m$) | 0.28 ($10^{-5}-10^{-3}m$) | 0.38-0.43 ($10^{-5}-10^{-3}m$) | 0.26 ($10^{-5}-10^{-3}m$) |
| $\alpha$ | – | 0.38 | – | 0.35 | 0.97 | – |
| $E_d^{L_e} = \frac{1}{2}Gb^3$, eV | 3,93 | 1,28 | 1,96 | 3,72 | 3,33 | 3,57 |
| $m_0 \cdot \alpha^2$ | 3.66-5.11 | 2.57 | 2.28 | 1.11 | 5.83-7.47 | 3.69 |
| $d_0$, nm | 23.6 | 14.4 | 13.6 | 22.6 | 23.8 | 28.0 |
| $\Delta\sigma_m$, GPa | 2.29-2.69 | 1.31 | 0.83 | 1.18 | 1.58-1.79 | 0.99 |
| $d_{\Sigma dis0}$, nm — $d_{GBs}$ / $d_{Ps}$ | 23.6 | 14.4 | 13.6 | 22.6 | 23.8 | 28.0 |
| $d_{\Sigma dis0}$, nm — $d_{GBg}$ / $d_{Pg}$ | 23.6 | 14.4 | 13.6 | 22.6 | 23.8 | 28.0 |
| $\bar{d}_{GB}, \bar{d}_P$ | ~135  24.0 | ~80  14.4 | ~80  13.6 | ~130  22 | ~150  24.8 | 150  28 |
| $\Delta\sigma_{\Sigma dis m}$, GPa — $-\sigma_{Pms}$ | 2.17  0.06 | 1.28  0.03 | 0.80  0.02 | 1.14  0.02 | 1.50  0.04 | 0.95  0.03 |
| $-\sigma_{Pmg}$ | 1.76  0.38 | 1.03  0.19 | 0.65  0.10 | 0.93  0.15 | 1.22  0.21 | 0.77  0.13 |
| $-\bar{\sigma}_{Pm}$ | 1.18  1.19 | 0.73  0.64 | 0.42  0.42 | 0.70  0.61 | 0.99  0.83 | 0.65  0.55 |

The values of $k$ at $\varepsilon = 0.002$ are taken, e.g., for α-Fe, Cu, Ni [6,7], Al [10], Zr, α-Ti [2] within the grain range enclosed in the frames.

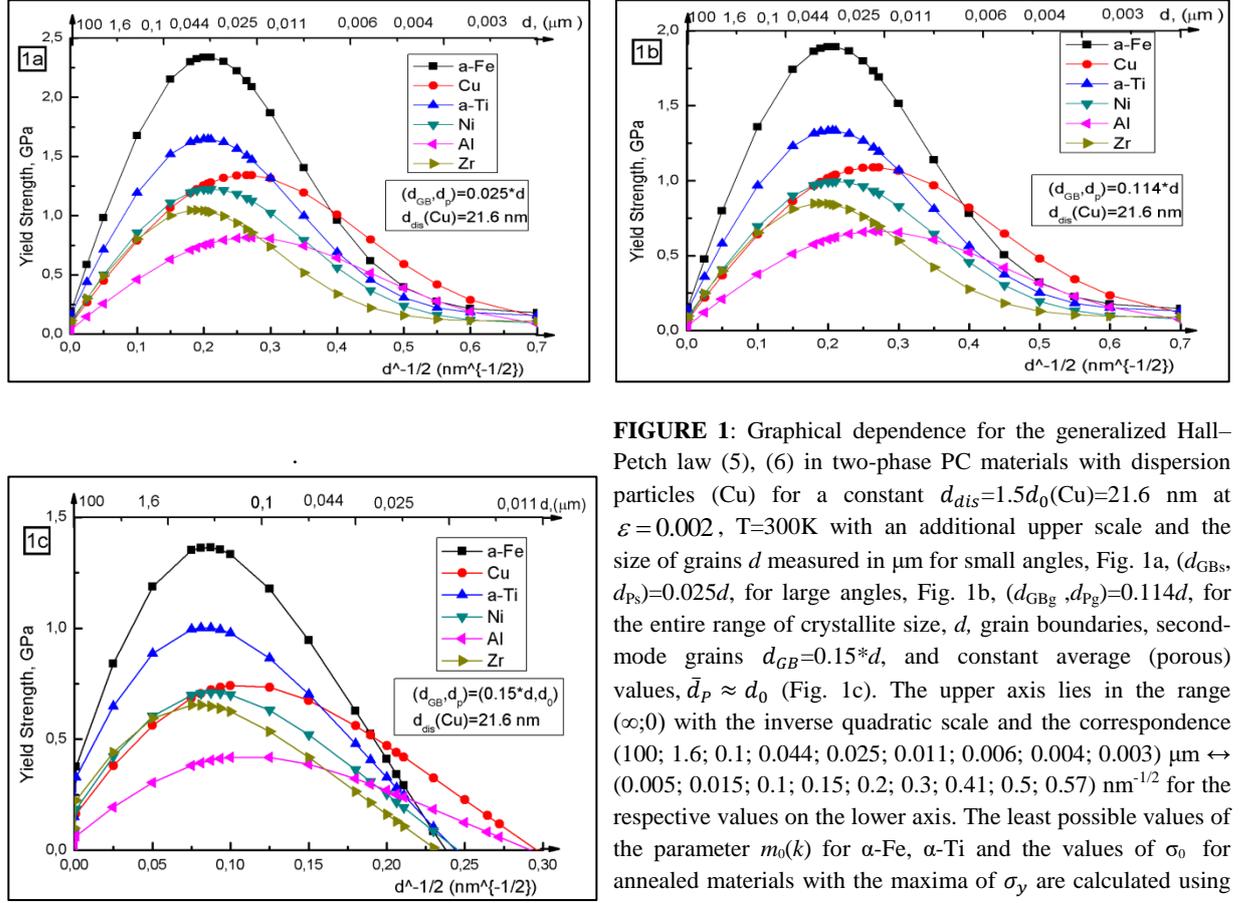

**FIGURE 1**: Graphical dependence for the generalized Hall–Petch law (5), (6) in two-phase PC materials with dispersion particles (Cu) for a constant $d_{dis}=1.5d_0(Cu)=21.6$ nm at $\varepsilon = 0.002$, T=300K with an additional upper scale and the size of grains $d$ measured in μm for small angles, Fig. 1a, ($d_{GBs}$, $d_{Ps}$)=0.025$d$, for large angles, Fig. 1b, ($d_{GBg}$, $d_{Pg}$)=0.114$d$, for the entire range of crystallite size, $d$, grain boundaries, second-mode grains $d_{GB}$=0.15*$d$, and constant average (porous) values, $\bar{d}_P \approx d_0$ (Fig. 1c). The upper axis lies in the range (∞;0) with the inverse quadratic scale and the correspondence (100; 1.6; 0.1; 0.044; 0.025; 0.011; 0.006; 0.004; 0.003) μm ↔ (0.005; 0.015; 0.1; 0.15; 0.2; 0.3; 0.41; 0.5; 0.57) nm$^{-1/2}$ for the respective values on the lower axis. The least possible values of the parameter $m_0(k)$ for α-Fe, α-Ti and the values of $\sigma_0$ for annealed materials with the maxima of $\sigma_y$ are calculated using the respective (see Table 1) extreme grain sizes $d_0$.

## TEMPERATURE DEPENDENCE OF YIELD STRENGTH AND EXTREME GRAIN SIZES FOR Al

We continue the study of TDE predicted in the one-phase model [2, 3] and the closely-packed two-phase model [4] of single-mode PC aggregates, now with third-phase (Cu) particles being present according to the three-phase model with $\sigma_{\Sigma dis}(\varepsilon)$ subject to (5), (6). The increase in temperature causes the value of $G(T)$ (and that of $\sigma_0(T)$) to decrease, whereas the linear parameters $b$, $d$ increase, with the same linear coefficient of the temperature expansion $\alpha_d$ [6] (for BCC and FCC materials, also see [10]), then the extreme grain size $d_{\Sigma 0}(\varepsilon,T)$ is shifted to the region of smaller grains, $d_{\Sigma 0}(\varepsilon,T) > d_{\Sigma 0}(\varepsilon, T')$ for $T > T'$ (in the same material phase [1,2]), according to

$$d_0(\varepsilon, T') = b_\varepsilon(T') \frac{\frac{1}{2}G(T')[b_\varepsilon(T')]^3}{1{,}59363 \cdot kT} = d_0(\varepsilon, T)g(\alpha_G, \alpha_d, T, T'),$$

$$g(\alpha_G, \alpha_d, T, T') = \left(\frac{b_\varepsilon(T')}{b_\varepsilon(T)}\right)^4 \frac{G(T')T}{G(T)T'} = (1 + \alpha_d(T' - T))^4 (1 - \alpha_G(T' - T))\frac{T}{T'}, \quad (7)$$

$$\text{for } b_\varepsilon(T') = b_\varepsilon(T)(1 + \alpha_d(T' - T)); \quad G(T') = (1 - \alpha_G(T' - T))G(T)$$

now separately for Al and Cu as non-coherent particles (for a constant $d_{dis}$), with average linear temperature coefficients of the expansion $\alpha_d$ and shear modulus $\alpha_G$, e.g., for Al, Cu, with $\alpha_d(Al) \in [19.5|_{T=150K} - 23.7|_{T=350K}], 10^{-6}K^{-1}$, $\alpha_d(Cu) \in [16.0|_{T=200K} - 17.0|_{T=350K}] \cdot 10^{-6}K^{-1}$ and $\alpha_G(Al; Cu) = (5.20; 4.54) \cdot 10^{-4}K^{-1}$ being approximately constant in the temperature range $[150K, 350K]$ (for Cu, see [11]). It follows from (7) that for $T$ varying in a small range the value of $d_0(\varepsilon,T)$ changes multiplicatively with the factor $g(\alpha_G, \alpha_d, T, T')$ and a correction due to Cu-particles. The behavior of both $\sigma_{\Sigma dis}(\varepsilon)$ and $\sigma_{\Sigma dis\ m}(\varepsilon)$ under a monotonous change of temperature is composed of the $T$-behavior of crystal and dislocation substructures of crystalline, weak and third (Cu) phases of samples according to the theoretical prescription (5)–(7). The results of our theoretical research for the $T$-dependence in the HP law for Al samples at large-angle GB, both without a third phase (given from [4] on the Fig 2a, for comparison) and with Cu-particles and a constant $T$-dependent size $d_{dis}(Cu)=21.6$ nm at T=300K, with the weight $U_{dis}$=0.05 for the same large-angle GB, as well as for a constant size of pores at the 2nd phase, (i.e., single-mode PC samples) in the entire range of diameters $d$ of the 1st-phase crystallites given by Table 2 and Fig. 2.

**TABLE 2:** The values $G$, $\sigma_0$, $d_0$, $\Delta\sigma_m$ for PC Al samples with $d_0$, $b$, $G$ for $T$=300 K are taken from Table 1, Fig. 1 and Ref. [2] at $\varepsilon$=0.002 and $d_{\Sigma dis0}$, $\Delta\sigma_{\Sigma dism}=(\sigma_{\Sigma dism}-\sigma_0)$ obtained from Fig. 2 for $d_{GB}=0$, $d_{Pg}=0.25*d$, which corresponds to the average weights of the phases $0.95*(1-nb/d; (n-n)b/d; nb/d)=(0.76; 0;0.19)$, $U_{dis}=0.05$, with no account for the weak phase grains with large-angle GB and constant (if any) size of porous $\bar{d}_P$ PC Al samples. The existence boundary $d_{LB}$ is estimated at 11 nm.

| Al,Cu  T,K | $G$(Al;Cu), GPa | $\sigma_0$(Al;Cu),MPa | $d_0$(Al;Cu), nm | $(\sigma_m-\sigma_0)$ Al, GPa | $d_{\Sigma dis0}$, nm; $(d_{Pg}=0,25d)$ | $\Delta\sigma_{\Sigma dism}$, GPa | $d_{\Sigma dis0}$, nm; $(\bar{d}_P=13{,}6 nm)$ | $\Delta\sigma_{\Sigma dism}$, GPa |
|---|---|---|---|---|---|---|---|---|
| 350 | 25.8; 43.0 | 21;6.8 | 11.7; 12.3 | 0.85 | 8.7 | 0.67 | 70.0 | 0.347 |
| 300 | 26.5; 44.0 | 22;7.0 | 13.6; 14.4 | 0.81 | 11.3 | 0.63 | 70.0 | 0.357 |
| 250 | 27.4; 45.0 | 23;7.16 | 16.3; 17.3 | 0.74 | 13.0 | 0.59 | 70.0 | 0.368 |
| 200 | 28.1; 46.0 | 23.5;7.3 | 20.4; 21.6 | 0.67 | 16.0 | 0.53 | 70.0 | 0.373 |
| 150 | 28.8; 47.0 | 24;7.5 | 27.2; 28.8 | 0.59 | 22.5 | 0.47 | 70.0 | 0.387 |

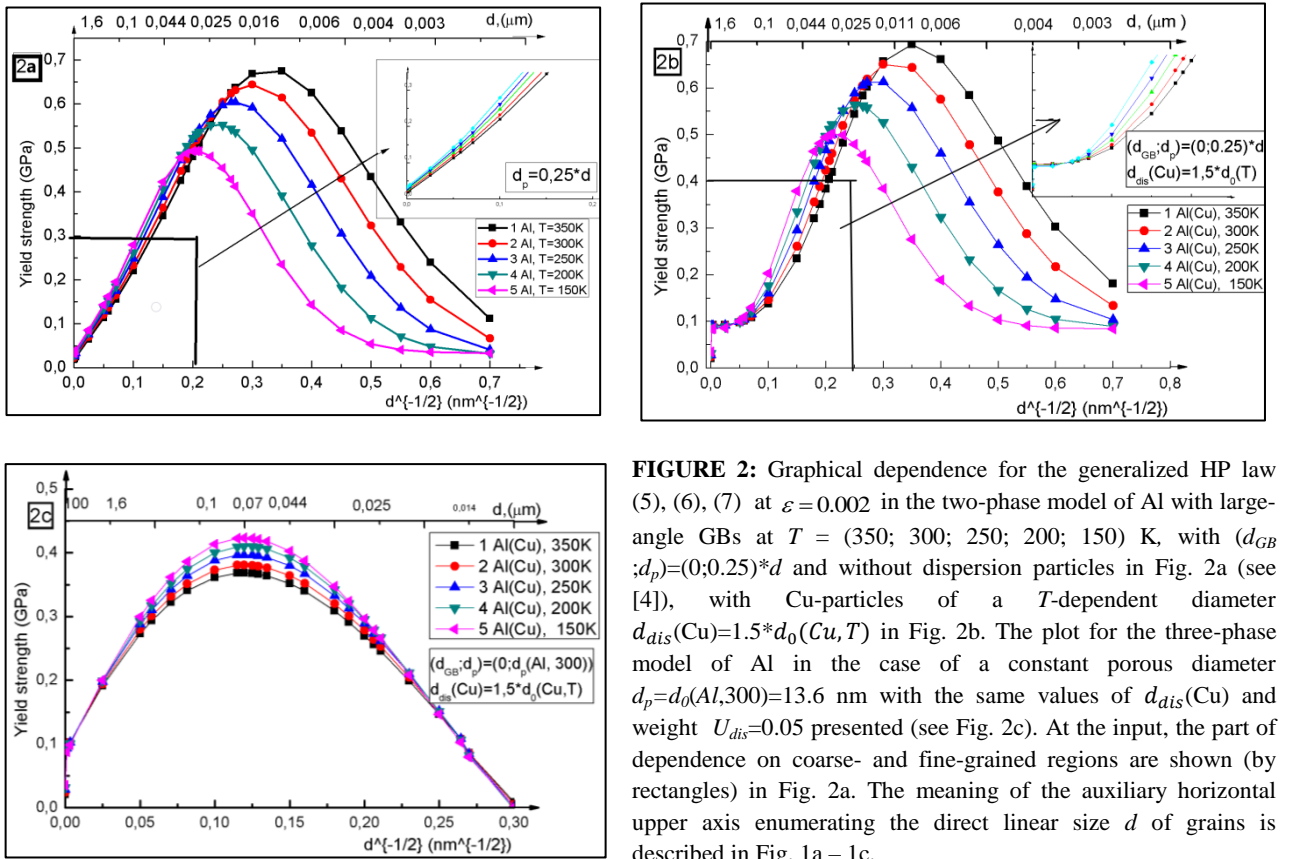

**FIGURE 2:** Graphical dependence for the generalized HP law (5), (6), (7) at $\varepsilon=0.002$ in the two-phase model of Al with large-angle GBs at $T=(350; 300; 250; 200; 150)$ K, with $(d_{GB};d_p)=(0;0.25)*d$ and without dispersion particles in Fig. 2a (see [4]), with Cu-particles of a $T$-dependent diameter $d_{dis}(Cu)=1.5*d_0(Cu,T)$ in Fig. 2b. The plot for the three-phase model of Al in the case of a constant porous diameter $d_p=d_0(Al,300)=13.6$ nm with the same values of $d_{dis}(Cu)$ and weight $U_{dis}=0.05$ presented (see Fig. 2c). At the input, the part of dependence on coarse- and fine-grained regions are shown (by rectangles) in Fig. 2a. The meaning of the auxiliary horizontal upper axis enumerating the direct linear size $d$ of grains is described in Fig. 1a – 1c.

## SUMMARY

From Table 1 and Fig. 1, the data for the Hall–Petch law at $T$=300 K in heterogeneous two-mode PC samples of α-Fe, Cu, Al, Ni, α-Ti, Zr for the two-phase model with dispersion (third-phase) particles of another compound from a PC composite metallic aggregate described by (5), (6), it follows that the extreme size $d_{\Sigma dis0}$ for the maximum $\Delta\sigma_{\Sigma dism}$ of yield strength decreases in comparison with $d_0$, $\Delta\sigma_m$ calculated according to (2), (3) in the one-phase approximation model with $d$-independent weights of the 1st and 2nd phases, $d_{\Sigma dis0}<d_0$, $\Delta\sigma_{\Sigma dism}<\Delta\sigma_m$. The doping of PC aggregates by non-coherent Cu-particles with the average diameter $d_{dis}(Cu)=1.5*d_0(Cu,300)$ near the extreme size of a one-mode polycrystalline Cu-sample leads to various kinds of dispersion response for samples within the entire range of first-phase grains, because the energy value of a unit dislocation, $E_d^{L_e}=\tfrac{1}{2}Gb^3$, for Cu is the least possible one among the other energy values. In the coarse- and (ultra-)fine-grained region of an entire PC aggregate, we observe dispersion hardening, whereas in the nano-crystalline region the situation is reversed. The extreme size of an average first-phase grain $d_{\Sigma dis0}$ may be both larger and smaller than that without dispersion particles, depending on the size $d_{dis}$ of particles and on their mechanical and physical nature. For small-angle GB, the difference $\Delta\sigma_{\Sigma dism}$ is larger than that for a large-angle GB, as in the case without a third phase. For a constant size of pores $d_P \sim d_0$, the weights of the 1st and 2nd phases become $d$-dependent, transforming PC samples (with a decrease in $d$) from PC aggregates with small-angle GB to those with large-angle GB, and reaching the impossibility of their existence in the NC region due to $\sigma_\Sigma<0$. The extreme size of grains for such samples is shifted (as in the case of a purely two-phase model) to a range of values readily accessible in experiment, $d_{\Sigma dis0}$ (α-Fe, Cu, Al, Ni, α-Ti, Zr)~(135;80;80;130;130;150) $nm$, starting from the SMC region with a decrease in $\Delta\sigma_{\Sigma dism}$ ($\leq$1GPa). An increase in the weight $U_{dis}$ of Cu-particles leads to a change of the above values. In turn, the study of temperature dependence for PC Al samples within the range of [150; 350] $K$ shows, based on Table 2 and Fig. 2, that the temperature-dimensional effect [2] discovered within the one-phase model approximation should be valid for samples with a $d$-independent small- and large-angle GB with dispersion hardening by Cu-particles of NC size, leading to an increase in $\Delta\sigma_{\Sigma m}$ and $\sigma_\Sigma$ for all $d<d_{\Sigma dis1}\sim 3d_{\Sigma dis0}$ [2] with an increase in temperature. For grains with $d>d_{\Sigma dis1}$, the behavior of $\sigma_{\Sigma dis}$ is opposite (and therefore usual for coarse- and fine-grained PC aggregates). The presence of the 2nd and 3rd phases (with $d$-independent weights for $U_{dis}$(Cu)=5%) diminishes the extreme grain size $d_{\Sigma dis0}$(T) in comparison with $d_0$(T). However, the 2nd phase with $d$-dependent weights of the 1st and 2nd phases with constant pores $d_p=d_0(300)=$ 13.6 nm leads to the disappearance of this effect in such PC samples, revealing an decrease in $\sigma_\Sigma$ with a growth of temperature in the entire range of grain size $d$, thus stabilizing the extreme size $d_{\Sigma dis0}$ in a PC Al aggregate at 70 nm. The situation is similar to a purely 2nd-phase model of PC Al aggregates without dispersion [4]. However, the presence of Cu-particles of T-dependent size, $d_{dis}$(Cu)=1.5*$d_0(Cu,T)$, for large-angle GB two-phase model of PC Al samples shows (as compared with pure 2-phase model) for hardening in the SMC, NC regions and un-hardening for CG, UFG regions of the Al first phase grains. Note, that the change of $d_{dis}$(Cu) to $\bar{d}_{dis}$(Cu)= $d_0(Cu,T)$ increases the values of $\Delta\sigma_{\Sigma dism}$ up to 0.1GPa for the same $U_{dis}$(Cu)=5%.

The present theoretical study demands an experimental verification of the predicted temperature-dimensional effect. Namely, one should verify the increase in the extreme size of grains $d_{\Sigma dis0}(\varepsilon,T)$ with a decreasing temperature and separately with a growth of accumulated strain $\varepsilon$ and with a simultaneous decrease, for a small-angle GB, of the maximum $\sigma_{\Sigma dis}(\varepsilon)$ in both single- and two-mode PC aggregates. The mentioned phenomena are characteristic of NC and SMC PC samples. The suggested theoretical model implies obvious perspectives of its application to new PC composite materials in aircraft and cosmic industry and has been tested experimentally using samples of the BT1-0 α-Ti alloy and ultrafine-grained PC samples [13]**.** The model allows one to obtain an analytic expression for the $\sigma$-$\varepsilon$ dependence of PC materials in the entire range of grain sizes, values of temperature and accumulated strain, correctly reflecting the experimental data. Further developments are expected for PC samples with various dispersion phases and multiple dislocation ensembles. This will permit a natural incorporation of twinning defects into the Taylor dislocation hardening mechanism due to their origin as combinations of different (among partial) dislocations, especially in the NC region.

## ACKNOWLEDGMENTS

The authors are grateful to the research fellows of ISPMS SB RAS for valuable discussions. The work is supported by the Program of Fundamental Research under the Russian Academy of Sciences for 2013 – 2020.